# Derivation and physical interpretation of the general solutions to the wave equations for electromagnetic potentials

Valerică Raicu

*Physics Department, University of Wisconsin-Milwaukee, Milwaukee, WI 53211, USA*

## Abstract

The inhomogeneous wave equations for the scalar, vector, and Hertz potentials are derived starting from retarded charge, current, and polarization densities and then solved in the reciprocal (or k-) space to obtain general solutions, which are formulated as nested integrals of such densities over the source volume, k-space, and time. The solutions thus obtained are inherently free of spatial singularities and do not require introduction by fiat of combinations of advanced and retarded terms as done previously to cure such singularities for the point-charge model. Physical implications of these general solutions are discussed in the context of specific examples involving either the real or reciprocal space forms of the different potentials. The present approach allows for k-space expansions for arbitrary distributions of charges and may lead to applications in fluorescence-based imaging and condensed matter research.

## 1. Introduction

Scalar and vector potentials – which are introduced as solutions to both time-independent (e.g., Poisson) and time-dependent (or wave) equations [1,2] – are extensively used in the study of virtually any important problem in the physical sciences and engineering, including condensed matter [3,4], inhomogeneous biological materials [5], antennas and electromagnetic propagation [6], and quantum optics and engineering [7], to name but a few areas. Methods exist for solving time-independent equations for distributions of charges and dipoles [8]. By contrast, methods for solving the inhomogeneous wave equations for potentials currently are restricted to special cases of point charges and dipoles [9]. While it is known that the general wave equations are satisfied separately by retarded and advanced solutions – with the latter being usually discarded based on causality considerations –, the interplay between the advanced and retarded solutions have remained open questions [1].

This report begins with straightforward derivations of the inhomogeneous wave equations for the electric scalar, magnetic vector, and electric Hertz potentials from retarded charge, current, and dipole densities, respectively, using the retarded calculus toolkit [10,11]. This approach circumvents the need to justify the choice of a specific Guage in deriving the wave equations from Maxwell's equations, helps avoid potential confusions regarding the definition of the quantities involved as functions of the "source" *or* the "field" coordinates, and allows for introduction of the equations of electrodynamics starting from potentials. General solutions of the wave equations are obtained as Fourier (or *k-form*) expansions of the charge, current, and polarization densities. Integration over both time and *k*-space leads to expressions that are in many ways similar to the known ones (as spatial integrals over charge, current, and polarization densities) but, at the same time, are free from singularities that usually have been removed, since Dirac's seminal work [12], through subtraction of the mean of advanced and retarded potentials. These results prompt some cautionary notes regarding the neglect of temporal derivatives to introduce the Poisson equation for static distributions of charge. This is because, even though a distribution of charge may be static, different parts of it are located at different positions in space causing their contributions to the potentials to present different degrees of retardation.

Finally, it shown herein that the k-space representation of the scalar, vector, and Hertz potentials derived from this theory could be useful in their own right, and not only as a way to obtain the real-space forms of the potentials. This point will be illustrated using a particular case of the Hertz potential of an oscillating point dipole which was proposed and employed by Cray, Shih, and Milonni (CSM) in their elegant treatment of the problem of radiation absorbed and radiated through stimulated emission based on purely classical arguments [13]. CSM proved that not only absorption but also stimulated emission may result in dipole oscillations at the same frequency as that of the exciting field. The ability to formulate k-space expansions of arbitrary distributions of charge afforded by the present approach will likely allow other interesting and useful physical problems to be tackled.

## 2. The wave equations for potentials

### 2.1. Derivation of the wave equations

Let us consider an arbitrary charge density, $\rho$, and a current density, $\vec{\jmath}$, located at a retarded position, that is, the position, denoted by *x', y', z'*, that the charge distribution or current had at the retarded time, $t' = t - R/c$, while *t* is the time at the evaluation (or field) point (see Fig. 2.1). For distributions of dipoles wherein the net charge is zero, a polarization density or simply polarization, $\vec{P}$, may be introduced instead of the charge density. The retarded charge, currents, and polarization densities are denoted by $[\rho] = \rho\left(\vec{r'}, t - \frac{|\vec{r}-\vec{r'}|}{c}\right)$, $[\vec{\jmath}] = \vec{\jmath}\left(\vec{r'}, t - \frac{|\vec{r}-\vec{r'}|}{c}\right)$, and $[\vec{P}] = P\left(\vec{r'}, t - \frac{|\vec{r}-\vec{r'}|}{c}\right)$, respectively. We are interested in computing the potentials and fields generated at the time *t* at a location denoted by unprimed spatial coordinates $x$, $y$, and $z$. The location of the retarded densities is represented by the position vector $\vec{r}' = (x', y', z')$ pointing to the direction where the distribution of charge was at the time $t'$, while the location of the potentials and fields at the time $t$ is represented by $\vec{r} = (x, y, z)$. Therefore, the vector $\vec{R} = \vec{r} - \vec{r}'$ of magnitude $R = |\vec{r} - \vec{r}'| = \sqrt{(x-x')^2 + (y-y')^2 + (z-z')^2}$ denotes the distance from the source to the field point. Note that the differential operators used (including spatial and temporal derivatives) can also be defined relative to both the primed and unprimed coordinates, and both types can be applied to the retarded and non-retarded physical quantities (see Appendix A).

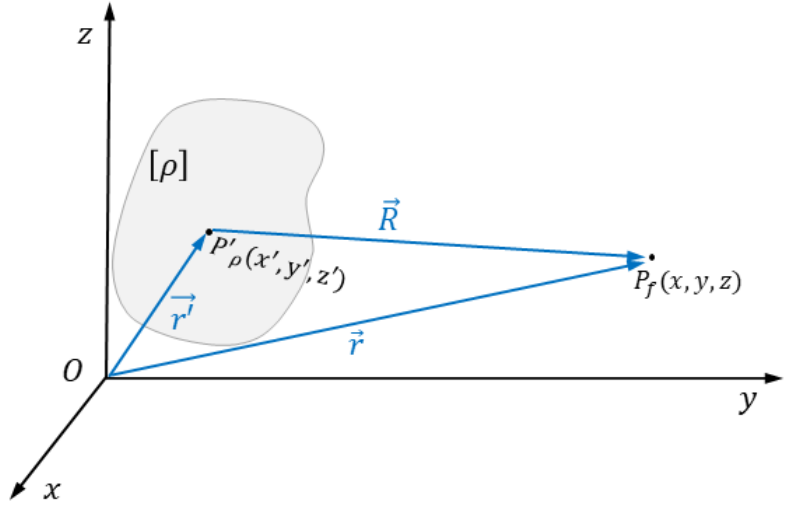

**Figure 2.1.** Illustration of the geometry used to calculate the potentials and fields at the field point ($P_f$) generated by distribution of charge $[\rho]$ whose instantaneous position is $P'_\rho$.

We will first derive an expression for the unprimed Laplacian of the ratio between $[\rho]$ and $R$, following a method previously applied by McQuistan to a general vector field [10], and then integrate the resulting expression over the volume centered around the charge distribution located at its retarded position to obtain the wave equation for the scalar potential. By applying the same method to the three different components of the current density and the polarization, respectively, and combining the resulting expressions into vector forms, we then obtain the wave equations for the magnetic and Hertz vector potentials.

To begin with, we use the vector identity (A7) in Appendix A to write

$$\vec{\nabla}\left\{\frac{[\rho]}{R}\right\} = \vec{\nabla}[\rho]\frac{1}{R} + [\rho]\vec{\nabla}\left(\frac{1}{R}\right) = -\frac{\hat{R}}{Rc}\frac{\partial[\rho]}{\partial t} + [\rho]\vec{\nabla}\left(\frac{1}{R}\right), \tag{2.1}$$

which, after applying $\vec{\nabla} \cdot$ to both of its sides and noticing that

$$\hat{R} = \vec{\nabla} R = -R^2 \vec{\nabla} \left( \frac{1}{R} \right),$$  (2.2)

becomes

$$\vec{\nabla} \cdot \vec{\nabla} \left\{ \frac{[\rho]}{R} \right\} = \left\{ \frac{\vec{\nabla} R}{c} \frac{\partial [\rho]}{\partial t} + \frac{R}{c} \frac{\partial}{\partial t} \vec{\nabla}[\rho] + \vec{\nabla}[\rho] \right\} \cdot \vec{\nabla} \left( \frac{1}{R} \right) + \left\{ \frac{R}{c} \frac{\partial [\rho]}{\partial t} + [\rho] \right\} \nabla^2 \left( \frac{1}{R} \right).$$

Using again equation (A7), and with the help of the well-known identity

$$\nabla^2 \left( \frac{1}{|\vec{r} - \vec{r}\prime|} \right) = -4\pi \delta^3 (\vec{r} - \vec{r}\prime),$$  (2.3)

where $\delta^3(\vec{r} - \vec{r}\prime)$ is the Dirac delta function, we obtain:

$$\vec{\nabla} \cdot \vec{\nabla} \left\{ \frac{[\rho]}{R} \right\} = \frac{1}{Rc^2} \frac{\partial^2 [\rho]}{\partial t^2} - \frac{R}{c} \frac{\partial [\rho]}{\partial t} 4\pi \delta^3 (\vec{r} - \vec{r}\prime) - [\rho] 4\pi \delta^3 (\vec{r} - \vec{r}\prime).$$  (2.4)

Since, by definition, the delta function is zero for any $\vec{r} \neq \vec{r}\prime$, while $R = |\vec{r} - \vec{r}\prime|$ is zero for $\vec{r} = \vec{r}\prime$, the product $R\delta^3(\vec{r} - \vec{r}\prime)$ is equal to zero for any $\vec{r}$, and thus the second term of equation (2.4) vanishes. Therefore,

$$\nabla^2 \left\{ \frac{[\rho]}{R} \right\} = \frac{1}{Rc^2} \frac{\partial^2 [\rho]}{\partial t^2} - 4\pi \delta^3 (\vec{r} - \vec{r}\prime)[\rho].$$  (2.5)

For vector functions, such as $[\vec{J}]$, the components $[J_x]$, $[J_y]$, and $[J_z]$ may be shown to satisfy similar equations which, multiplied by their respective unit vector and added together, give

$$\nabla^2 \left\{ \frac{[\vec{J}]}{R} \right\} = \frac{1}{Rc^2} \frac{\partial^2 [\vec{J}]}{\partial t^2} - 4\pi \delta^3 (\vec{r} - \vec{r}\prime)[\vec{J}].$$  (2.6)

Applying the same procedure to the components of another useful vector quantity, the electric polarization density (or, simply, polarization), $[\vec{P}]$, we can write the following expression:

$$\nabla^2 \left\{ \frac{[\vec{P}]}{R} \right\} = \frac{1}{Rc^2} \frac{\partial^2 [\vec{P}]}{\partial t^2} - 4\pi \delta^3 (\vec{r} - \vec{r}\prime)[\vec{P}].$$  (2.7)

Note that, unlike charge and current, polarization is usually defined as the volume density of dipoles, and thus the "polarization density" has the same meaning as "polarization."

Multiplying both sides of equation (2.5) by $\frac{1}{4\pi \varepsilon_v}$ and integrating with respect to $d^3 r\prime = dx\prime dy\prime dz\prime$ over entire space we obtain:

$$\nabla^2 \phi - \frac{1}{c^2} \frac{\partial^2}{\partial t^2} \phi = -\frac{1}{\varepsilon_v} \rho,$$  (2.8a)

where

$$\phi\left(\vec{r'}, t'\right) = \frac{1}{4\pi\varepsilon_v} \iiint \frac{[\rho]}{R} d^3 r' \tag{2.8b}$$

is the retarded scalar potential and

$$\rho(\vec{r}, t) \equiv \iiint_{-\infty}^{\infty} \rho\left(\vec{r'}, t - \frac{|\vec{r} - \vec{r'}|}{c}\right) \delta^3(\vec{r'} - \vec{r}) d^3 r' \tag{2.8c}$$

provides a prescription for computing the charge density as a function of the unprimed coordinates.

Similarly, by multiplying equation (2.6) by $\frac{\mu_v}{4\pi}$ and integrating, we have:

$$\nabla^2 \vec{A} - \frac{1}{c^2} \frac{\partial^2}{\partial t^2} \vec{A} = -\mu_v \vec{J}, \tag{2.9a}$$

where

$$\vec{A} = \frac{\mu_v}{4\pi} \iiint d^3 r' \frac{[\vec{J}]}{R} \tag{2.9b}$$

is the retarded vector potential and

$$\vec{J}(\vec{r}, t) \equiv \iiint_{-\infty}^{\infty} d^3 r' \vec{J}\left(\vec{r'}, t - \frac{|\vec{r} - \vec{r'}|}{c}\right) \delta^3(\vec{r'} - \vec{r}) \tag{2.9c}$$

is the current density at the potential (or field) position.

Finally, multiplying both sides of equation (2.7) by $\frac{1}{4\pi\varepsilon_v}$ and integrating with respect to $d^3 r' = dx' dy' dz'$, we obtain:

$$\nabla^2 \vec{\Pi} - \frac{1}{c^2} \frac{\partial^2}{\partial t^2} \vec{\Pi} = -\frac{\vec{P}}{\varepsilon_v}, \tag{2.10a}$$

where

$$\vec{\Pi} = \frac{1}{4\pi\varepsilon_v} \iiint d^3 r' \frac{[\vec{P}]}{R} \tag{2.10b}$$

is the retarded electric Hertz potential and

$$\vec{P}(\vec{r}, t) \equiv \iiint_{-\infty}^{\infty} d^3 r' \vec{P}\left(\vec{r'}, t - \frac{|\vec{r} - \vec{r'}|}{c}\right) \delta^3(\vec{r'} - \vec{r}) \tag{2.10c}$$

is the polarization at the coordinates of the potential. Similar formulas also may be written for magnetic polarization, which nevertheless are not of interest in the present work.

Interestingly, while none of the densities (of charge, current, or polarization) on the right-hand-sides of equations (2.8a), (2.9a), and (2.10a) are retarded quantities, the associated potentials are computed from their retarded counterparts. These observations may not be immediately obvious when the wave equations for the potentials are derived from Maxwell's equations using the Lorentz condition.

Similar wave equations also may be derived using "advanced" forms of the charge densities (i.e., charge densities evaluated at $t' = t + R/c$), implying that advanced potentials also can exist, in principle. This brings up the old dilemma as to whether the advanced solutions to the wave equations have physical meaning. We will address this question in the next section wherein we will obtain solutions to the wave equations for the general case of arbitrary distributions of charges and currents.

It is possible to relate the scalar and vector potentials to the Hertz potential [14]. If the polarization vector is defined such that

$$[\rho] = -\left[\vec{\nabla}' \cdot \vec{P}\right] \tag{2.11}$$

and

$$[\vec{J}] = \left[\frac{\partial \vec{P}}{\partial t}\right], \tag{2.12}$$

then, inserting (2.11) and (A15) into (2.8b) and assuming the charge to be concentrated within a finite region of space, we obtain

$$\phi = -\vec{\nabla} \cdot \vec{\Pi}, \tag{2.13}$$

while inserting (2.12) and (A5) in (2.9b), we have

$$\vec{A} = \frac{1}{c^2}\frac{\partial \vec{\Pi}}{\partial t}. \tag{2.14}$$

The last two relationships are useful in some applications (see section 4).

## 2.2. Relationships among potentials and between them and electromagnetic fields

So far, we have only inquired how the charge and current densities lead separately to scalar and vector potentials. The same mathematical apparatus used above, taken together with the continuity equation,

$$\left[\vec{\nabla}' \cdot \vec{J}\right] = -\left[\frac{\partial \rho}{\partial t}\right], \tag{2.15}$$

where the differential operators $\vec{\nabla}'$ and $\left[\frac{\partial \rho}{\partial t}\right]$ are taken with respect to the primed coordinates, allows us to determine how the scalar and vector potentials relate to one another. The derivation of the relationship between the two potentials, known as the *Lorenz's condition* [10,15], is extensively covered in the literature and also included in Appendix B for convenience, while here we only provide its result:

$$\vec{\nabla} \cdot \vec{A} = -\frac{1}{c^2}\frac{\partial}{\partial t}\phi. \tag{2.16}$$

In the standard electromagnetic theory, the Lorenz gauge condition embodied by this equation is chosen as a means to decouple a more complicated differential equation into two separate wave equations, one for the vector potential and one for the scalar potential; in the current theoretical framework, the same expression emerges from the continuity equation.

    The equations of electrodynamics introduced by Maxwell [16-19] using a molecular vortex model that is currently considered obsolete, were compactified by Heaviside using vector calculus notations [20] and subsequently regarded as mathematical postulates [21]. Long after the Maxwell-Heaviside equations were published, interest arose in deriving them from special relativity considerations [22], the continuity equation for localized scalar and vector sources [23], Coulomb's law together with the postulate of the constancy of the velocity of light [24], and, more recently, from the wave equations for the scalar and vector potentials, which were introduced through the continuity equation [25]. Having already derived the wave equations for the scalar and vector potentials (via a different route) as well as well as the Lorenz relation, we may now introduce both the usual relations between potentials and fields,

$$\vec{E} = -\vec{\nabla}\Phi - \frac{\partial}{\partial t}\vec{A}, \tag{2.17}$$

$$\vec{B} = \vec{\nabla} \times \vec{A}, \tag{2.18}$$

and Maxwell's equations following the approach suggested by Heras and Heras, as shown in Appendix C. It only remains to be emphasized here that the electromagnetic fields in Maxwell's equations are retarded quantities, in the same way and for the same reasons that the potentials are retarded, although the former are usually introduced starting from the unprimed coordinates. Indeed, the fields can be traced directly to charges and currents via Jefimenko's equations [1,11].

## 3. Solving the wave equations for potentials

As discussed in section 2, the relationships (2.8a), (2.9a), and (2.10a) are mathematical *identities* that can be satisfied by arbitrary functions $[\rho]$, $[\vec{J}]$, and $[\vec{P}]$ of the *retarded time and position*. At the same time, they are equations with respect to potentials, i.e., they admit unique solutions $\phi$, $\vec{A}$, and $\vec{\Pi}$ when the derivatives are taken with respect to the unprimed coordinates. The exact analytical solutions of the wave equations are usually guessed but not rigorously derived except for the particular case of point-charges, which is tackled using Green's function [9]. Here we will introduce a general method starting from the classical approach of Fourier transforming the original differential equations [26], solving them in the reciprocal (or k-space), and then returning to the real space via an inverse Fourier transform. A similar program was previously proposed by Hu [27], aiming to derive a solution to the scalar potential of the form given by Eqn. (2.8b). Unfortunately, implementation of that program has been marred by inconsistencies (such as in its use of primed vs. unprimed coordinates), and its main goal has not been fully achieved. In fact, as we will see below, the goal of any such program needs to be modified.

### 3.1 The scalar potential

Let us use the notation

$$f(\vec{r}, t) \equiv c^2 \frac{\rho(\vec{r}, t)}{\varepsilon_v}, \tag{3.1a}$$

with $\rho(\vec{r}, t)$ given by eqn. (2.8c) and, therefore,

$$f(\vec{r}, t) = \iiint_{-\infty}^{\infty} d^3 r' f\left(\vec{r'}, t - \frac{|\vec{r} - \vec{r'}|}{c}\right) \delta^3(\vec{r'} - \vec{r}), \tag{3.1b}$$

to rewrite equation (2.8a) as

$$\frac{\partial^2 \phi(\vec{r}, t)}{\partial t^2} - c^2 \nabla^2 \phi(\vec{r}, t) = f(\vec{r}, t). \tag{3.2}$$

Here we must recall that the potential is a function of the time and space coordinates at the field point (i.e., unprimed coordinates), in addition to being a function of the retarded charge density (which is expressed in the primed coordinates), while $\rho(\vec{r}, t)$ and, therefore, $f(\vec{r}, t)$, on the right-hand-side of this equation is a function of the unprimed coordinates.

As it is well known, if the condition

$$\iiint_{-\infty}^{\infty} d^3 r |\psi(\vec{r})| < \infty \tag{3.3}$$

is fulfilled by an arbitrary function $\psi(\vec{r})$, then such a function may be expressed as the Fourier transform of another function, $\Psi(\vec{k})$, and vice versa. It may be easily seen that this condition is satisfied by $f(\vec{r}, t)$ for arbitrary distributions of charge, subject only to the reasonable requirement that they be confined to finite regions of space. We can also safely assume that the condition (3.3) is satisfied for $\phi(\vec{r})$ at $-\infty$ and $+\infty$. We can only guess at this time that it should also behave well at the position of the charge when expressed in spherical coordinates, because the wave equations should formally admit a combination of advanced and retarded forms as their solutions, which should mitigate the overall behavior at zero. This of course may raise the objection that the advanced solution violates causality. However, we can proceed with our plan on purely mathematical grounds and revisit this point once the actual solution is obtained.

For $\phi(\vec{r}, t)$ and $f(\vec{r}, t)$ in equation (3.2), the forward and inverse Fourier transforms are

$$\Phi(\vec{k}, t) = \iiint_{-\infty}^{\infty} d^3 r\, \phi(\vec{r}, t) e^{-i\vec{k}\cdot\vec{r}}, \tag{3.4}$$

$$\phi(\vec{r}, t) = \frac{1}{(2\pi)^3} \iiint_{-\infty}^{\infty} d^3 k\, \Phi(\vec{k}, t) e^{i\vec{k}\cdot\vec{r}}, \tag{3.5}$$

and, respectively,

$$F(\vec{k}, t) = \iiint_{-\infty}^{\infty} d^3 r\, f(\vec{r}, t) e^{-i\vec{k}\cdot\vec{r}}, \tag{3.6}$$

$$f(\vec{r}, t) = \frac{1}{(2\pi)^3} \iiint_{-\infty}^{\infty} d^3 k\, F(\vec{k}, t) e^{i\vec{k}\cdot\vec{r}}, \tag{3.7}$$

where $\vec{k} \equiv (k_x, k_y, k_z)$. Note that, while Eqn. (3.6) is the standard form of the Fourier transform, we will find it useful to change the integration variable in that expression from $r$ to $r'$ by inserting identity (3.1b) into (3.6). Thus,

$$F(\vec{k}, t) = \iiint_{-\infty}^{\infty} d^3 r' \left[ \iiint_{-\infty}^{\infty} d^3 r\, f\left(\vec{r'}, t - \frac{|\vec{r} - \vec{r'}|}{c}\right) e^{-i\vec{k}\cdot\vec{r}} \delta^3(\vec{r} - \vec{r'}) \right] = \iiint_{-\infty}^{\infty} d^3 r'\, f\left(\vec{r'}, t\right) e^{-i\vec{k}\cdot\vec{r'}}. \tag{3.6'}$$

Inserting equations (3.5) and (3.7) into the wave equation (3.2), and noticing that

$$\nabla^2 \phi(\vec{r}, t) = -\frac{1}{(2\pi)^3} \iiint_{-\infty}^{\infty} d^3 k\, k^2 \Phi(\vec{k}, t) e^{i\vec{k}\cdot\vec{r}},$$

and

$$\frac{\partial^2 \phi(\vec{r}, t)}{\partial t^2} = \frac{1}{(2\pi)^3} \iiint_{-\infty}^{\infty} d^3 k\, \frac{\partial^2 \Phi(\vec{k}, t)}{\partial t^2} e^{i\vec{k}\cdot\vec{r}},$$

we obtain

$$\iiint_{-\infty}^{\infty} d^3 k \left[ \frac{\partial^2 \Phi(\vec{k}, t)}{\partial t^2} + (ck)^2 \Phi(\vec{k}, t) - F(\vec{k}, t) \right] e^{i\vec{k}\cdot\vec{r}} = 0, \tag{3.8}$$

where $k^2 = k_x{}^2 + k_y{}^2 + k_z{}^2$ was used. For the inverse Fourier transform given by equation (3.8) to be identically zero, the expression between the square brackets must itself be identically zero. Thus, we have:

$$\frac{\partial^2 \Phi(\vec{k},t)}{\partial t^2} + (ck)^2 \Phi(\vec{k},t) = F(\vec{k},t). \tag{3.9}$$

The following particular solution of equation (3.9) may be constructed using standard methods:

$$\Phi(\vec{k},t) = \frac{1}{2ick} e^{ickt} \int_{-\infty}^{t} d\tau \, e^{-ick\tau} F(\vec{k},\tau) - \frac{1}{2ick} e^{-ickt} \int_{-\infty}^{t} d\tau \, e^{ick\tau} F(\vec{k},\tau), \tag{3.10}$$

in which the time variable $t$ under the integrals is the dummy variable $\tau$.

A general solution to equation (3.9) may be written as the sum between the particular solution given by Eqn. (3.10) and the solution to the associated homogenous equation, i.e.,

$$\Phi(\vec{k},t) = \zeta_1(\vec{k}) e^{ickt} + \zeta_2(\vec{k}) e^{-ickt} + \frac{1}{2ick} \int_{-\infty}^{t} d\tau \, e^{ick(t-\tau)} F(\vec{k},\tau) - \frac{1}{2ick} \int_{-\infty}^{t} d\tau \, e^{-ick(t-\tau)} F(\vec{k},\tau),$$

where $\zeta_1(\vec{k})$ and $\zeta_1(\vec{k})$ are time-independent quantities that may be determined from the requirement that the general solution satisfies the boundary conditions. Substituting $F(\vec{k},\tau)$ from identity (3.6') into the last equation and the resulting expression into equation (3.5), we obtain the general expression for the scalar potential in terms of $f\left(\vec{r'},t\right)$ and, thus, $\rho\left(\vec{r'},t\right)$, as a function of the position of the charge distribution and the time at the position of the potentials and fields,

$$\phi(\vec{r},t) = \frac{1}{(2\pi)^3} \iiint_{-\infty}^{\infty} d^3k \left[ \zeta_1(\vec{k}) e^{ickt} + \zeta_2(\vec{k}) e^{-ickt} \right] e^{i\vec{k}\cdot\vec{r}} +$$

$$\frac{1}{(2\pi)^3} \iiint_{-\infty}^{\infty} d^3k \int_{-\infty}^{t} d\tau \iiint_{-\infty}^{\infty} d^3r' f\left(\vec{r'},\tau\right) \frac{1}{2ick} \left[ e^{ick(t-\tau)} - e^{-ick(t-\tau)} \right] e^{i\vec{k}\cdot(\vec{r}-\vec{r'})}. \tag{3.11}$$

Note that, were we to use identity (3.6) instead of (3.6'), the exponential $e^{i\vec{k}\cdot(\vec{r}-\vec{r'})}$ in equation (3.11) would have become equal to one, leading to different solutions. The same would have happened if $f(\vec{r},t)$ in equation (3.2) were incorrectly written in terms of the primed coordinates (i.e., as a function of the retarded charge), in which case, the integration variable and the function variable also would have been different. This clearly illustrates the necessity of good coordinates bookkeeping.

To determine the integration constants in Eqn. (3.11), we set the potential $\phi$ and its temporal derivative to zero for $R \equiv \left|\vec{r} - \vec{r'}\right| \to \infty$. Under these conditions, and after carrying out one or more of the integrations in the second term of Eqn. (3.11) resulting in, e.g., Eqn. (3.14) below, it may be easily seen that the second term and its derivative become zero as well; this leads to $\zeta_1(\vec{k}) = \zeta_2(\vec{k}) = 0$. With these boundary conditions and after changing the order of integration and replacing the exponentials by a sine function, the last equation for the scalar potential becomes:

$$\phi(\vec{r}, t) = \frac{1}{8\pi^3 c} \int_{-\infty}^{t} d\tau \iiint_{-\infty}^{\infty} d^3 r' f\left(\vec{r'}, \tau\right) \iiint_{-\infty}^{\infty} d^3 k \frac{1}{k} \sin[ck(t - \tau)] \, e^{i\vec{k} \cdot (\vec{r} - \vec{r'})}. \tag{3.12}$$

To carry out the triple integral with respect to $k$ in the second term of Eqn. (3.12), we switch to spherical coordinates and replace $\vec{k} \cdot \vec{R}$ by $kR\cos\theta$. Thus, we can write for that integral,

$$\iiint_{-\infty}^{\infty} d^3 k \frac{1}{k} \sin[ck(t - \tau)] \, e^{i\vec{k} \cdot \vec{R}} = \int_{0}^{2\pi} d\varphi \int_{0}^{\pi} d\theta \sin\theta \int_{0}^{\infty} dk \sin[ck(t - \tau)] \, e^{ikR\cos\theta} k,$$

which, after using the substitutions $u = \cos\theta$ and $du = -\sin\theta \, d\theta$ and integrating, becomes

$$4\pi \int_{0}^{\infty} dk \frac{1}{R} \sin[ck(t - \tau)] \sin(kR).$$

Inserting this integral into Eqn. (3.12), we obtain

$$\phi(\vec{r}, t) = \frac{1}{2\pi^2 c} \int_{-\infty}^{t} d\tau \iiint_{-\infty}^{\infty} d^3 r' f\left(\vec{r'}, \tau\right) \int_{0}^{\infty} dk \frac{1}{R} \sin[ck(t - \tau)] \sin(kR). \tag{3.13}$$

After using (3.1a) and $\sin a \sin b = \frac{1}{2}[\cos(a - b) - \cos(a + b)]$, we obtain

$$\phi(\vec{r}, t) = \frac{c}{4\pi^2 \varepsilon_v} \int_{-\infty}^{t} d\tau \iiint_{-\infty}^{\infty} d^3 r' \frac{1}{R} \rho\left(\vec{r'}, \tau\right) \int_{0}^{\infty} dk \cos\left[ck\left(t - \tau - \frac{R}{c}\right)\right] -$$

$$\frac{c}{4\pi^2 \varepsilon_v} \int_{-\infty}^{t} d\tau \iiint_{-\infty}^{\infty} d^3 r' \frac{1}{R} \rho\left(\vec{r'}, \tau\right) \int_{0}^{\infty} dk \cos\left[ck\left(t - \tau + \frac{R}{c}\right)\right],$$

which, since we may write $\int_{0}^{\infty} dk \cos(kx) = \pi\delta(x)$, is equivalent to

$$\phi(\vec{r}, t) = \frac{c}{4\pi\varepsilon_v} \int_{-\infty}^{t} d\tau \iiint_{-\infty}^{\infty} d^3 r' \frac{1}{R} \rho\left(\vec{r'}, \tau\right) \delta\left[c\left(t - \tau - \frac{R}{c}\right)\right] - \frac{c}{4\pi\varepsilon_v} \int_{-\infty}^{t} d\tau \iiint_{-\infty}^{\infty} d^3 r' \frac{1}{R} \rho\left(\vec{r'}, \tau\right) \delta\left[c\left(t - \tau + \frac{R}{c}\right)\right].$$

Using well-known properties of the Dirac delta function, this equation may be re-written as

$$\phi(\vec{r}, t) = \frac{1}{4\pi\varepsilon_v} \int_{-\infty}^{t} d\tau \iiint_{-\infty}^{\infty} d^3 r' \frac{1}{R} \rho\left(\vec{r'}, \tau\right) \delta\left[\tau - \left(t - \frac{R}{c}\right)\right] - \frac{1}{4\pi\varepsilon_v} \int_{-\infty}^{t} d\tau \iiint_{-\infty}^{\infty} d^3 r' \frac{1}{R} \rho\left(\vec{r'}, \tau\right) \delta\left[\tau - \left(t + \frac{R}{c}\right)\right], \tag{3.14}$$

where $R \equiv \left|\vec{r} - \vec{r'}\right|$ (see Fig. 2.1). This expression, like all of its preceding forms, is an exact solution to the wave equation for the scalar potential, which appears to behave as $\phi \to 0$ for $R \to 0$ (or $\vec{r} \to \vec{r'}$), since the asymptotic behavior for $R \to 0$ of the delta functions in both integrals is the same: $\delta(\tau - t)$. This somewhat surprising result nevertheless confirms the assumption of integrability of the potential (3.3) made from the outset.

Equation (3.14) also may be reduced to a more familiar form by changing the order of integration and using the sifting property of the delta function. In this case the first integral over $\tau$ gives $f\left(\vec{r'}, t - \frac{R}{c}\right)$, as $t - \frac{R}{c}$ falls within the interval $(-\infty, t]$ for any R, while the second integral is equal to zero over most of that interval since $t + \frac{R}{c}$ falls outside of it for any $R$ except $R = 0$. Thus, after returning to the notation given by Eqn. (3.1), we can rewrite the last result as

$$\phi(\vec{r}, t) = \frac{1}{4\pi\varepsilon_v} \iiint_{-\infty}^{\infty} d^3r' \frac{\rho\left(\vec{r'}, t - \frac{R}{c}\right)}{R}, \text{ for } R > 0. \tag{3.15}$$

which is the same as equation (2.8b) and is the generally accepted expression for the potential of a distribution of charges [1,28]. However, when written in this form, it obfuscates the behavior of the potential for $R \to 0$.

The complete expression, which is valid for any positive $R$, is

$$\phi(\vec{r}, t) = \frac{1}{4\pi\varepsilon_v} \iiint_{-\infty}^{\infty} d^3r' \frac{\rho\left(\vec{r'}, t - \frac{R}{c}\right)}{R} - \frac{1}{4\pi\varepsilon_v} \delta_\rho(R), \tag{3.16}$$

where $\delta_\rho(\vec{R})$ is a *delta-like function* defined as

$$\delta_X(R) = \begin{cases} 0, \ R > 0 \\ \lim_{R \to 0} \iiint_{-\infty}^{\infty} d^3r' \frac{X\left(\vec{r'}, t - \frac{R}{c}\right)}{R}, \ R = 0 \end{cases}, \tag{3.17}$$

in which $X = \rho$ for the case of Eqn. (3.16). The reason for designating $\delta_X(R)$ as a "delta-like function" is not that it has the same properties as Dirac's delta function, but simply that it is infinite at and centered around $\vec{R} = 0$ and zero for $\vec{R} \neq 0$ (where $\vec{R}$ takes both positive and negative values). For $R \to 0$, the first integral in Eqn. (3.16) takes the same form as (3.17) and therefore we have, successively,

$$\phi\left(\vec{r} = \vec{r'}, t\right) = \frac{1}{4\pi\varepsilon_v} \lim_{R \to 0} \iiint_{-\infty}^{\infty} d^3r' \frac{\rho\left(\vec{r'}, t - \frac{R}{c}\right)}{R} - \frac{1}{4\pi\varepsilon_v} \delta_\rho(R) = \frac{1}{4\pi\varepsilon_v} \delta_\rho(R) - \frac{1}{4\pi\varepsilon_v} \delta_\rho(R) = 0,$$

which differs from the result $\phi\left(\vec{r} = \vec{r'}, t\right) = \infty$ that is usually assumed. We will return to this result in the Discussion section.

## 3.2. The vector potential

By separating the wave equation (2.9a) for the vector potential into three scalar equations written in terms of the spatial components $A_w$ and $g_w$ (with $w = x, y, z$), Fourier transforming the three scalar equations, and recombining their respective solutions into vector form, we obtain:

$$\vec{A}(\vec{r}, t) = \frac{1}{(2\pi)^3} \iiint_{-\infty}^{\infty} d^3k \left[ \vec{\eta_1}(\vec{k}) e^{ikct} + \vec{\eta_2}(\vec{k}) e^{-ikct} \right] e^{i\vec{k}\cdot\vec{r}} +$$

$$\frac{1}{(2\pi)^3} \iiint_{-\infty}^{\infty} d^3k \int_{-\infty}^{t} d\tau \iiint_{-\infty}^{\infty} d^3r' \frac{1}{2ikc} \vec{g}\left(\vec{r'}, \tau\right) \left[ e^{ikc(t-\tau)} - e^{-ikc(t-\tau)} \right] e^{i\vec{k}\cdot(\vec{r}-\vec{r'})}, \tag{3.18a}$$

where

$$\vec{g}(\vec{r}, t) \equiv c^2 \mu_v \vec{j}(\vec{r}, \tau). \tag{3.18b}$$

Here again, we chose the boundary conditions such that $\vec{\eta_1} = \vec{\eta_2} = 0$ and then change the order of integration to obtain:

$$\vec{A}(\vec{r}, t) = \frac{c\mu_v}{8\pi^3} \int_{-\infty}^{t} d\tau \iiint_{-\infty}^{\infty} d^3r' \vec{j}\left(\vec{r'}, \tau\right) \iiint_{-\infty}^{\infty} d^3k \frac{1}{k} \sin[ck(t-\tau)] e^{i\vec{k}\cdot(\vec{r}-\vec{r'})}, \tag{3.19}$$

and, further,

$$\vec{A}(\vec{r}, t) = \frac{\mu_v}{4\pi} \int_{-\infty}^{t} d\tau \iiint_{-\infty}^{\infty} d^3r' \frac{1}{R} \vec{j}\left(\vec{r'}, \tau\right) \delta\left[\tau - \left(t - \frac{R}{c}\right)\right] - \frac{\mu_v}{4\pi} \int_{-\infty}^{t} d\tau \iiint_{-\infty}^{\infty} d^3r' \frac{1}{R} \vec{j}\left(\vec{r'}, \tau\right) \delta\left[\tau - \left(t + \frac{R}{c}\right)\right]. \tag{3.20}$$

This is an exact solution to the wave equation for the scalar potential, which, similarly to the scalar potential, behaves asymptotically as $\phi \to 0$ for $R \to 0$. Using the same reasoning as employed in the derivation of equation (3.1), we may also write,

$$\vec{A}(\vec{r}, t) = \frac{\mu_v}{4\pi} \iiint_{-\infty}^{\infty} d^3r' \frac{\vec{j}\left(\vec{r'}, t - \frac{R}{c}\right)}{R}, \text{ for } R > 0, \tag{3.21}$$

which is identical to Eqn. (2.9b) and is subject to the same limitations as the "classical" scalar potential. The complete expression, which is also valid for $R = 0$, is

$$\vec{A}(\vec{r}, t) = \frac{\mu_v}{4\pi} \iiint_{-\infty}^{\infty} d^3r' \frac{\vec{j}\left(\vec{r'}, t - \frac{R}{c}\right)}{R} - \frac{\mu_v}{4\pi} \delta_{\vec{j}}(R), \tag{3.22}$$

where $\delta_{\vec{j}}(\vec{R})$ is the delta-like function introduced through Eqn. (3.17), with $X = \vec{j}$. As with the scalar potential, the vector potential expressed by Eqn. (3.22) becomes zero, and not infinite, for $R = 0$.

### 3.3. The Hertz potential

By separating the wave equation (2.10a) for the vector potential into three scalar equations written in terms of the spatial components $\Pi_w$ and $g_w(\vec{r}, t)$ (with $w = x, y, z$), Fourier transforming the three scalar equations, and recombining their respective solutions into vector form, here too we obtain:

$$\vec{\Pi}(\vec{r}, t) = \frac{1}{(2\pi)^3} \iiint_{-\infty}^{\infty} d^3k \left[ \overrightarrow{\chi_1}(\vec{k}) e^{ikct} + \overrightarrow{\chi_2}(\vec{k}) e^{-ikct} \right] e^{i\vec{k}\cdot\vec{r}} +$$

$$\frac{1}{(2\pi)^3} \iiint_{-\infty}^{\infty} d^3k \int_{-\infty}^{t} d\tau \iiint_{-\infty}^{\infty} d^3r' \frac{1}{2ikc} \vec{h}\left(\overrightarrow{r'}, \tau\right) \left[ e^{ikc(t-\tau)} - e^{-ikc(t-\tau)} \right] e^{i\vec{k}\cdot(\vec{r}-\overrightarrow{r'})}, \tag{3.23a}$$

where

$$\vec{h}(\vec{r}, \tau) \equiv c^2 \frac{\vec{P}(\vec{r}, \tau)}{\varepsilon_v}, \tag{3.23b}$$

and $\vec{P}(\vec{r}, t)$ is the electrical polarization vector.

With our now familiar choice of boundary conditions, for which $\overrightarrow{\chi_1} = \overrightarrow{\chi_2} = 0$, and after changing the order of integration and combining the temporal exponentials, we obtain:

$$\vec{\Pi}(\vec{r}, t) = \frac{c}{8\pi^3 \varepsilon_v} \int_{-\infty}^{t} d\tau \iiint_{-\infty}^{\infty} d^3r' \vec{P}(\vec{r}, \tau) \iiint_{-\infty}^{\infty} d^3k \frac{1}{k} \sin[ck(t-\tau)] \, e^{i\vec{k}\cdot(\vec{r}-\overrightarrow{r'})}. \tag{3.24}$$

Similarly to the other two potentials, we may rewrite (4.25) successively as

$$\vec{\Pi}(\vec{r}, t) = \frac{1}{4\pi\varepsilon_v} \int_{-\infty}^{t} d\tau \iiint_{-\infty}^{\infty} d^3r' \frac{1}{R} \vec{P}(\vec{r}, \tau) \delta\left[\tau - \left(t - \frac{R}{c}\right)\right] - \frac{1}{4\pi\varepsilon_v} \int_{-\infty}^{t} d\tau \iiint_{-\infty}^{\infty} d^3r' \frac{1}{R} \vec{P}(\vec{r}, \tau) \delta\left[\tau - \left(t + \frac{R}{c}\right)\right], \tag{3.25}$$

and

$$\vec{\Pi}(\vec{r}, t) = \frac{1}{4\pi\varepsilon_v} \iiint_{-\infty}^{\infty} d^3r' \frac{\vec{P}\left(\overrightarrow{r'}, t - \frac{R}{c}\right)}{R}, \text{ for } R > 0. \tag{3.26}$$

This equation is identical to Eqn. (2.10b) and is subject to the same constraints as the ones corresponding to the scalar and vector potentials. The complete expression, which is also valid for $R = 0$, is

$$\vec{\Pi}(\vec{r}, t) = \frac{1}{4\pi\varepsilon_v} \iiint_{-\infty}^{\infty} d^3r' \frac{\vec{P}\left(\overrightarrow{r'}, t - \frac{R}{c}\right)}{R} - \frac{1}{4\pi\varepsilon_v} \delta_{\vec{P}}(R), \tag{3.27}$$

where $\delta_{\vec{P}}(R)$ is the delta-like function introduced through Eqn. (3.17), with $X = \vec{P}$, which implies that the Hertz potential is zero, and not infinite, for $R = 0$.

### 3.4. Particular cases

We next consider two particular cases, that of a point-charge, $q$, and that of an oscillating point dipole, $\vec{\mu}(t)$, with instantaneous positions denoted by $\overrightarrow{r'_c}(t)$ and $\overrightarrow{r'_d}(t)$, respectively (see Fig. 3.1). These entities are assumed to move in arbitrary directions with instantaneous velocities $\vec{v_c}$ and $\vec{v_d}$. For the point-charge, the charge and current densities may be written as

$$\rho\left(\vec{r'}, t\right) = q\,\delta^3\left[\vec{r'} - \overrightarrow{r'_c}(t)\right], \tag{3.28}$$

and

$$\vec{J}\left(\vec{r'}, t\right) = q\vec{v_c}(t)\delta^3\left[\vec{r'} - \overrightarrow{r'_c}(t)\right], \tag{3.29}$$

respectively, while for the point dipole, the polarization density is

$$\vec{P}\left(\vec{r'}, t\right) = \vec{\mu}(t)\delta^3\left[\vec{r'} - \overrightarrow{r'_d}(t)\right], \tag{3.30}$$

where $\vec{\mu}(t)$ is the dipole moment.

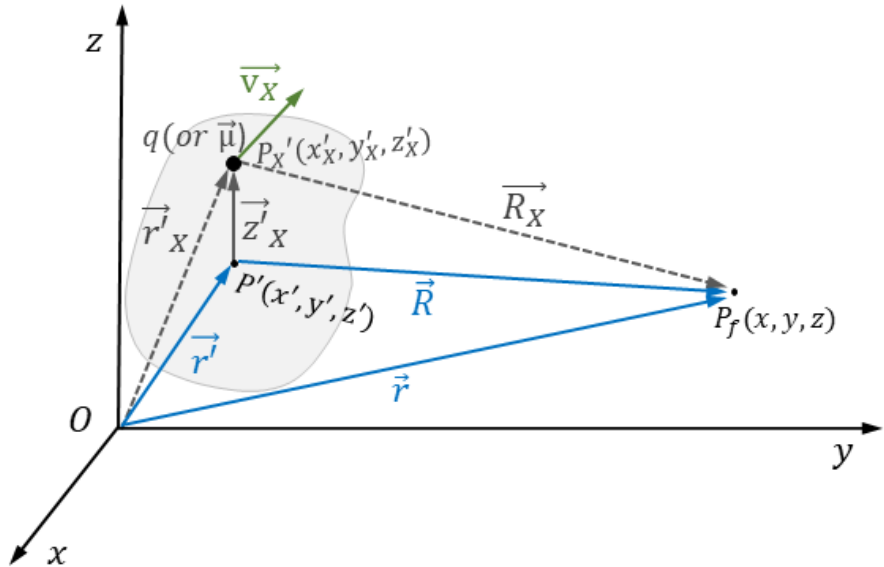

**Figure D1.** Illustration of the geometry used to calculate the potentials and fields at the field point ($P_f$) generated by a distribution of charge consisting either of a point charge $q$ or point dipole (with moment $\vec{\mu}$) whose instantaneous positions are $P'_X$ (with $X = c$ or $d$) and are moving in the $z$-direction with instantaneous velocity $\overrightarrow{v_X}$ relative to $P'_X$. It is obvious from the figure that the following relation holds: $\overrightarrow{R_X}(\tau) = \vec{r} - \overrightarrow{r'_X}(\tau)$.

Inserting the charge, current and polarization density in Eqns. (3.14), (3.20), and (3.25), changing the integration variable so that it becomes the same as the argument of the delta function [29], and integrating over time gives by carefully keeping track of the different times involved (see Appendix D), we obtain:

$$\phi = \frac{q}{4\pi\varepsilon_v \left|\overrightarrow{R_c}\left(t - \frac{R_c}{c}\right)\right|} \frac{1}{1 - \widehat{R_c}\left(t - \frac{R_c}{c}\right)\cdot\overrightarrow{\beta_c}\left(t - \frac{R_c}{c}\right)} - \frac{q}{4\pi\varepsilon_v}\delta_{R_c}, \tag{3.31}$$

$$\vec{A} = \frac{q\mu_v}{4\pi\left|\overrightarrow{R_c}\left(t - \frac{R_c}{c}\right)\right|} \frac{c\overrightarrow{\beta_c}\left(t - \frac{R_c}{c}\right)}{1 - \widehat{R_c}\left(t - \frac{R_c}{c}\right)\cdot\overrightarrow{\beta_c}\left(t - \frac{R_c}{c}\right)} - \frac{q\mu_v c\overrightarrow{\beta_c}(t)}{4\pi}\delta_{R_c}, \tag{3.32}$$

$$\overrightarrow{\Pi} = \frac{1}{4\pi\varepsilon_v \left|\overrightarrow{R_d}\left(t - \frac{R_c}{c}\right)\right|} \frac{\vec{\mu}\left(t - \frac{R_c}{c}\right)}{1 - \widehat{R_d}\left(t - \frac{R_d}{c}\right)\cdot\overrightarrow{\beta_d}\left(t - \frac{R_c}{c}\right)} - \frac{\vec{\mu}(t)}{4\pi\varepsilon_v}\delta_{R_d} \tag{3.33}$$

where $\delta_{R_X}$ is a *delta-like function* defined as

$$\delta_{R_X} = \begin{cases} & 0, \ R_X > 0 \\ \lim_{R_X \to 0} \frac{1}{\left|\overrightarrow{R_X}\left(t + \frac{R_X}{c}\right)\right|} \frac{1}{1 - \frac{\overrightarrow{R_X}\left(t + \frac{R_X}{c}\right)}{\left|\overrightarrow{R_X}\left(t + \frac{R_X}{c}\right)\right|}\cdot\frac{\vec{v}\left(t + \frac{R_X}{c}\right)}{c}} = \infty, \ R_X = 0, \end{cases} \tag{3.34}$$

$t' = t - R/c$, $\vec{R_X} = \vec{r} - \overrightarrow{r'}_X$ (with "$X$" being either "c" or "d"), and the other symbols have their usual meaning. The first two expressions are similar to, yet significantly different from, the relativistic Liénard-Wiechert potentials [29] due to the presence of their second terms, which are compactly represented by delta-like functions contributed by the advanced term in the expression for the potential in the reciprocal space. In all three expressions, the delta function terms are only used as shorthand notations for the fact that the second integrals in Eqns. (3.14), (3.20), and (3.25), from which these terms were derived, are zero everywhere except at $R_X = 0$, where they become infinite.

## 4. Discussion

The solutions to the inhomogeneous wave equations for the scalar, vector, and Hertz potentials obtained above, when integrated over the entire $k$-space as well as time, produced for $R > 0$ relations given by Eqns. (3.15), (3.21), and (3.26) that that are identical to the expressions introduced in section 2. Those are the standard solutions found in textbooks [1,28] for the three potentials. However, when expanded to include $R \to 0$, the expressions for the potentials, given by Eqns. (3.16), (3.22), and (3.27), each comprises not one but two different terms: the usual retarded term and an advanced term whose only effect is to ensure that the potential vanishes in the limit $R \to 0$. This unanticipated result can be traced back to the retarded, $\delta\left[\tau - \left(t - \frac{R}{c}\right)\right]$, and advanced, $\delta\left[\tau - \left(t + \frac{R}{c}\right)\right]$, delta functions in Eqns. (3.14), (3.20), and (3.25), which exhibit the same asymptotic behavior, $\sim\delta(\tau - t)$, but are preceded by opposite signs; thus, each pair of terms in the aforementioned equations vanishes, thereby rendering the potentials equal to zero for $R = 0$, and not infinity as is usually believed. Therefore, the advanced forms do contribute to the final solutions to the wave equations, but without violating causality. This result obtained from a purely classical electrodynamics approach reconciles the point-particle model of the electron with the quantum mechanics demand [12] that the potentials (as well as the fields) do not present singularities. High-energy particle collision experiments or (low-energy) experiments with two-dimensional electron gas phases [30] may provide inspiration for testing this previously unrealized prediction of electrodynamics.

To fully understand the results of the present theoretical framework, an important question remains to be asked: Does the scalar potential obtained in this work reduce itself to the solution to the well-known Poisson equation for a static distribution of charge? The Poisson equation,

$$c^2\nabla^2\phi(\vec{r}, t) = -f(\vec{r}, t), \tag{4.1}$$

is easily obtained by setting the temporal derivative in Eqn. (3.2) to zero. By Fourier-transforming it by following the same procedure as above, we obtain the simple algebraic form

$$(ck)^2\Phi(\vec{k}) = F(\vec{k}),$$

which has the solution

$$\Phi(\vec{k}) = \frac{F(\vec{k}, \tau)}{c^2 k^2} = \frac{1}{\varepsilon_v k^2}\iiint_{-\infty}^{\infty} d^3 r' \rho\left(\vec{r'}\right) e^{-i\vec{k}\cdot\vec{r'}}.$$

Upon return to real space, this last result becomes

$$\phi(\vec{r}, t) = \frac{1}{(2\pi)^3 \varepsilon_v}\iiint_{-\infty}^{\infty} d^3 r' \frac{\rho(\vec{r'})}{R}\iiint_{-\infty}^{\infty} d^3 k\, \frac{1}{k^2} e^{i\vec{k}\cdot\vec{R}},$$

or, after switching to spherical coordinates in k,

$$\phi(\vec{r}, t) = \frac{4\pi}{(2\pi)^3 \varepsilon_v}\iiint_{-\infty}^{\infty} d^3 r' \frac{\rho(\vec{r'})}{R}\int_0^{\infty} d(kR)\,\frac{\sin(kR)}{kR} = \frac{1}{4\pi\varepsilon_v}\iiint_{-\infty}^{\infty} d^3 r' \frac{\rho(\vec{r'})}{R}. \tag{4.2}$$

Note that Eqn. (4.2), while identical to the classical one, is not a particular case of Eqn. (3.16)! This is because, even though a distribution of charge may be static, its temporal derivative is not necessarily zero, since different parts of the charge distribution are located at different positions in space and, therefore, the fields and potentials associated with those parts need to start propagating at different times in order to arrive simultaneously at the position (denoted by the unprimed coordinates) where they are evaluated. That remains true even for point-charges (or dipoles), similar to retardation playing a role in integrating over distributions of charges in classical electrodynamics [1,31]. Through this effect, time enters the wave equations and causes the temporal derivative in Eqn. (3.2) to differ from zero even for stationary charges. In other words, one can formulate the wave equation but not the Poisson equation when retardation is considered; this requires that Eqn. (4.2) must contain an additional term, $-(1/4\pi\varepsilon_v)\,\delta_\rho(R)$. However, while this effect is important for "charge-centric" problems, i.e., for cases in which the charge distribution is known and the potential (or field) outside of it must be determined, it does not cause difficulties for "field-centric" problems wherein the field is generated by some charge distribution located outside the region of interest, and the questions asked concern the behavior of the material and its contribution to the potentials and fields in the region of interest. In this latter case, one may formulate and solve the Poisson or Laplace equation following standard, though sometimes laborious, approaches [5,32].

The existence of delta functions embedded in the real-space forms of the potentials, which can be traced back to the exponentials $e^{i\vec{k}\cdot(\vec{r}-\overrightarrow{r'})}$ in the Fourier-expanded versions of the potentials, pose no formal or conceptual difficulty when handled within the k space and, moreover, they also facilitate the analysis of some interesting practical problems. To illustrate this last statement, let us write the k-space expansion of $\vec{\Pi}$ expressed by Eqn. (3.24) for the case of a point-dipole placed at a fixed position, for which the polarization is given by Eqn. (3.30). We thus write

$$\vec{\Pi}(\vec{r},t) = \frac{c}{8\pi^3\varepsilon_v}\iiint_{-\infty}^{\infty}d^3k\,\frac{1}{k}\int_{-\infty}^{t}d\tau\,\vec{\mu}(\tau)\sin[kc(t-\tau)-\vec{k}\cdot\overrightarrow{R_d}], \tag{4.3}$$

where we used the fact that integrals containing $\sin[\vec{k}\cdot\overrightarrow{R_d}(\tau)]$ (over the entire k-space) may be shown to vanish when switching to spherical coordinates. Equation (4.1) is identical to the electric Hertz potential expression used by Cray, Shih, and Milonni [13] to compute the electric field emitted by a point dipole subjected to a sinusoidal excitation, which was chosen as a suitable model to study absorption and stimulated emission by an atom. In so doing, CSM followed [33] the more common route based on the Green's function for point particles [9]. Within the present theoretical framework, the fields may be derived from the scalar and vector potentials via the usual relations given by equations (2.13) and (2.14) which, in turn, may be derived from Eqn. (4.3) using (2.17). In particular, the electric field is (see Appendix E):

$$\vec{E}_d(\vec{r},t) = -\frac{c}{8\pi^3\varepsilon_v}\iiint_{-\infty}^{\infty}d^3k\,\vec{k}[\hat{k}\cdot\vec{\mu}(\tau)]\sin[kc(t-\tau)-\vec{k}\cdot\overrightarrow{R_d}] +$$

$$\frac{c}{8\pi^3\varepsilon_v}\iiint_{-\infty}^{\infty}d^3k\int_{-\infty}^{t}d\tau\,k\,\vec{\mu}(\tau)\sin[kc(t-\tau)-\vec{k}\cdot\overrightarrow{R_d}] - \frac{1}{8\pi^3\varepsilon_v}\vec{\mu}(t)\iiint_{-\infty}^{\infty}d^3k\,\cos(\vec{k}\cdot\overrightarrow{R_d}), \tag{4.4}$$

which, strictly speaking, differs from that of CSM by a third term which, nevertheless, does not impact the results if the field is computed for times after radiation emission is complete, i.e., for "free" fields (in which case, $\vec{\mu}(t)$]. The energy associated with the interference between the incident field, $\vec{E}_i(\vec{r}, t) = \hat{x} E_{i0} \sin(\omega_i t - \vec{k}_i \cdot \vec{R})$, and the field radiated by the dipole, given by Eqn. (4.4), is

$$W = \varepsilon_v \int d^3 R \ \vec{E}_i(\vec{r}, t) \cdot \vec{E}_d(\vec{r}, t),$$

where the incident field and the dipole are both considered oriented in the $\hat{x}$ direction. Integration over the entire space leads to several delta functions of the type $\delta^3(\vec{k} - \vec{k}_i)$, which, upon integration over $\vec{k}$, have the effect of setting $\vec{k}$ equal to $\vec{k}_i$ and the angular frequency $\omega_k$ to $\omega_i$ in the rate of change,

$$\frac{\partial}{\partial t} W = -\omega_i E_{i0} \mu(t) \sin \omega_i t,$$

of the energy radiated (or absorbed) by the dipole [13]. This remarkable result obtained from classical electrodynamics, which indicates that the stimulated radiation occurs at the same frequency as that of the stimulating field (regardless of its initial frequency), has inspired the work presented in the present report, whose author is interested in understanding absorption, spontaneous and stimulated emission, as well as energy transfer between fluorescent molecules [34] and using it in micro-spectroscopy experiments. With the ability to formulate k-space expansions of "tailor-made" distributions of charge (including in molecules and molecular assemblies) afforded by the present work, the k-form representation of the potentials will be instrumental in tackling those as well as other important practical problems.

## Acknowledgements

I wish to thank my colleague, Prof. Daniel Agterberg for very helpful suggestions regarding mathematical techniques and subtleties, and Prof. Peter Milonni for kindly providing detailed answers to my many questions regarding some of his publications. This work has been partly supported through a Discovery and Innovation Grant from the University of Wisconsin-Milwaukee (DIG 101X453).

## Competing interests statement

The author declares that there are no competing financial interests.

## Supplemental Materials

### Appendix A: Derivation of useful retardation calculus equations

In this paper we denote the "retarded" position that the charge distribution or currents had at a previous time, $t' = t - R/c$, by primed spatial coordinates, $x'$, $y'$, $z'$, and the position at which the fields are computed at the present time $t$, by the unprimed spatial coordinates $x$, $y$, and $z$. The locations of the retarded charge and currents, $[\rho] = \rho\left(\vec{r'}, t - \frac{|\vec{r}-\vec{r'}|}{c}\right)$ and $[\vec{j}] = \vec{j}\left(\vec{r'}, t - \frac{|\vec{r}-\vec{r'}|}{c}\right)$, are therefore represented by the position vector $\vec{r'} = (x', y', z')$ where the distribution of charge was at the retarded time $t' = t - R/c$, and the location of the fields at the present time, $t$, by $\vec{r} = (x, y, z)$. The vector $\vec{R} = \vec{r} - \vec{r'}$, of magnitude

$$R = |\vec{r} - \vec{r'}| = \sqrt{(x-x')^2 + (y-y')^2 + (z-z')^2}, \tag{A1a}$$

is the distance between the sources and the point of measurement of the field, and the unit vector is given by

$$\hat{R} = \frac{\vec{R}}{R}. \tag{A1b}$$

Integrals over charges and currents are performed either over an element of volume at the location of the charge, $dV'$, or the location of the field, $dV$.

Whenever we perform operations with the del operator with respect to the unprimed coordinates, we use $\vec{\nabla}$, while for operations with respect to the prime coordinates, we use $\vec{\nabla}'$. Since $\vec{r} - \vec{r'} = -(\vec{r'} - \vec{r})$, it may be immediately seen that

$$\vec{\nabla}\left(\frac{1}{|\vec{r}-\vec{r'}|}\right) = -\vec{\nabla}'\left(\frac{1}{|\vec{r}-\vec{r'}|}\right) = -\frac{\vec{R}}{R^2}. \tag{A2}$$

In general, for a scalar or vector quantity that is a function of $x'$, $y'$, $z'$, and $t'$, we may write:

$$\frac{\partial[X]}{\partial t'} \equiv \frac{\partial[X]}{\partial(t-R/c)} = \left[\frac{\partial X}{\partial t}\right], \tag{A3}$$

which means that the derivative of a retarded quantity with respect to the retarded time, $t'$, gives the same result as taking the derivative of the unretarded quantity (which is a function of $x$, $y$, $z$, and $t$) with respect to the present time followed by replacement of the spatial and temporal coordinates with their retarded values. Alternatively, since $\partial t' = \partial t$, also one may write:

$$\frac{\partial [X]}{\partial t'} \equiv \frac{\partial [X]}{\partial (t - R/c)} = \frac{\partial [X]}{\partial t}. \tag{A4}$$

Combining the last two expressions, we have

$$\frac{\partial [X]}{\partial t} = \left[\frac{\partial X}{\partial t}\right]. \tag{A5}$$

Next, we shall derive useful relationships between the spatial and temporal derivatives of the retarded charge and current densities. The derivative of a retarded scalar quantity, $[X]$, with respect to the unprimed coordinate $x$, may be written, using the chain rule, in terms of the derivative with respect to the retarded time, $t'$, as

$$\hat{x}\frac{\partial [X]}{\partial x} = \hat{x}\frac{\partial [X]}{\partial (t - R/c)}\bigg|_{x',y',z'}\frac{\partial (t - R/c)}{\partial x},$$

$$\hat{y}\frac{\partial [X]}{\partial x} = \hat{y}\frac{\partial [X]}{\partial (t - R/c)}\bigg|_{x',y',z'}\frac{\partial (t - R/c)}{\partial y},$$

$$\hat{z}\frac{\partial [X]}{\partial x} = \hat{z}\frac{\partial [X]}{\partial (t - R/c)}\bigg|_{x',y',z'}\frac{\partial (t - R/c)}{\partial z},$$

where all partial derivatives with respect to the spatial coordinates $x$, $y$, and $z$ were taken as zero (because the retarded function is independent of the unprimed coordinates). Since

$$t' = t - \frac{R}{c} = t - \frac{1}{c}\sqrt{(x - x')^2 + (y - y')^2 + (z - z')^2},$$

and, hence,

$$\frac{\partial (t - R/c)}{\partial x} = -\frac{x - x'}{cR},$$

$$\frac{\partial (t - R/c)}{\partial y} = -\frac{y - y'}{cR},$$

$$\frac{\partial (t - R/c)}{\partial z} = -\frac{z - z'}{cR},$$

we get

$$\hat{x}\frac{\partial [X]}{\partial x} = -\hat{x}\frac{x - x'}{cR}\frac{\partial [X]}{\partial t}. \tag{A6a}$$

$$\hat{y}\frac{\partial [X]}{\partial y} = -\hat{y}\frac{y - y'}{cR}\frac{\partial [X]}{\partial t}, \tag{A6b}$$

$$\hat{\mathbf{z}}\frac{\partial[X]}{\partial z} = -\hat{\mathbf{z}}\frac{z-z'}{cR}\frac{\partial[X]}{\partial t}, \tag{A6c}$$

where we also used identity (A4).

Adding together relations (A6) and using identity (A1b), we obtain immediately:

$$\vec{\nabla}[X] = -\frac{\hat{R}}{c}\frac{\partial[X]}{\partial t}. \tag{A7}$$

In general, we may write for the operator $\vec{\nabla}$ applied to a retarded function (of $x'$, $y'$, $z'$, and $t'$)

$$\vec{\nabla} = -\frac{\hat{R}}{c}\frac{\partial}{\partial t}. \tag{A8}$$

This operator also may be applied via a scalar product to a retarded vector function, such as

$$[\vec{X}] = [X_x]\hat{\mathbf{x}} + [X_y]\hat{\mathbf{y}} + [X_z]\hat{\mathbf{z}},$$

to obtain:

$$\vec{\nabla}\cdot[\vec{X}] = -\frac{\hat{R}}{c}\cdot\frac{\partial[\vec{X}]}{\partial t}. \tag{A9}$$

We may also use (A8) to write the directed derivative,

$$\hat{R}\cdot\vec{\nabla} = -\frac{1}{c}\frac{\partial}{\partial t}. \tag{A10}$$

Following the same arguments as above, one may write for the spatial derivatives with respect to the primed coordinates,

$$\hat{\mathbf{x}}\frac{\partial[X]}{\partial x'} = \hat{\mathbf{x}}\frac{\partial[X]}{\partial x'}\bigg|_{y',z',t'} - \hat{\mathbf{x}}\frac{x-x'}{cR}\frac{\partial[X]}{\partial t}\bigg|_{x',y',z'}, \tag{A11a}$$

$$\hat{\mathbf{y}}\frac{\partial[X]}{\partial y'} = \hat{\mathbf{y}}\frac{\partial[X]}{\partial y'}\bigg|_{x',z',t'} - \hat{\mathbf{y}}\frac{y-y'}{cR}\frac{\partial[X]}{\partial t}\bigg|_{x',y',z'}, \tag{A11b}$$

$$\hat{\mathbf{z}}\frac{\partial[X]}{\partial z'} = \hat{\mathbf{z}}\frac{\partial[X]}{\partial z'}\bigg|_{x',y',t'} - \hat{\mathbf{z}}\frac{z-z'}{cR}\frac{\partial[X]}{\partial t}\bigg|_{x',y',z'}, \tag{A11c}$$

where all the partial derivatives of the spatial coordinates $x'$, $y'$, $z'$ with respect to each other were taken as zero since they are independent of one another. By summing up equations (A11) we obtain

$$\vec{\nabla}'[X] = \left[\vec{\nabla}'X\right] - \frac{\hat{R}}{c}\frac{\partial[X]}{\partial t}. \tag{A12}$$

Similarly, we may also write

$$\overrightarrow{\nabla'} \cdot \left[\vec{X}\right] = \left[\overrightarrow{\nabla'} \cdot \vec{X}\right] - \frac{\hat{R}}{c} \cdot \frac{\partial[\vec{X}]}{\partial t}. \tag{A13}$$

By combining the above equations with one another and using well-known vector identities, other useful identities are rather easily obtained (Jefimenko 1989). For instance, by eliminating the terms that include the derivative with respect of $t$ from equations (A9) and (A13), we get:

$$\left[\overrightarrow{\nabla'} \cdot \vec{X}\right] = \overrightarrow{\nabla} \cdot \left[\vec{X}\right] + \overrightarrow{\nabla'} \cdot \left[\vec{X}\right], \tag{A14}$$

while by dividing both sides of (A14) by $R$, adding and subtracting $\frac{\vec{R}}{R^2}$ to and from its right-hand-side, and then using (A2), we obtain:

$$\frac{\left[\overrightarrow{\nabla'} \cdot \vec{X}\right]}{R} = \overrightarrow{\nabla} \cdot \left\{\frac{[\vec{X}]}{R}\right\} + \overrightarrow{\nabla'} \cdot \left\{\frac{[\vec{X}]}{R}\right\}. \tag{A15}$$

## Appendix B: Derivation of the Lorenz condition

By taking the derivative with respect to time of equation (2.8b) and using identity (A3), we may write successively:

$$-\mu_v \varepsilon_v \frac{\partial}{\partial t} \phi = -\frac{\mu_v}{4\pi} \int \frac{\partial}{\partial t} \left\{ \frac{[\rho]}{R} \right\} dV' = -\frac{\mu_v}{4\pi} \int \frac{1}{R} \left[ \frac{\partial \rho}{\partial t} \right] dV'. \tag{B1}$$

In the last step we also used the fact that $R$ is independent of time, as it is a property of space and not of the charge distribution. Substituting $\left[ \frac{\partial \rho}{\partial t} \right]$ from the continuity equation (2.15) and transforming the resulting expression with the help of identity (A17) for $[\vec{X}] = [\vec{J}]$, we have

$$-\mu_v \varepsilon_v \frac{\partial}{\partial t} \phi = \frac{\mu_v}{4\pi} \int \vec{\nabla} \cdot \left\{ \frac{[\vec{J}]}{R} \right\} dV' + \frac{\mu_v}{4\pi} \int \vec{\nabla'} \cdot \left\{ \frac{[\vec{J}]}{R} \right\} dV'. \tag{B2}$$

Using Gauss's theorem, we may convert the second integral to a surface integral. Its integrand then approaches zero because the surface of integration encloses the entire space (i.e., $R \to \infty$) (Jefimenko 1989), which means that the ratio between a finite current and an infinite radius is zero. With this, equation (B2) becomes

$$-\mu_v \varepsilon_v \frac{\partial}{\partial t} \phi = \vec{\nabla} \cdot \left\{ \frac{\mu_v}{4\pi} \int \frac{[\vec{J}]}{R} dV' \right\}, \tag{B3}$$

where the entire expression in the curly brackets may be recognized as the vector potential, $\vec{A}$, which is defined by equation (2.9b). Thus, we obtain Lorenz's condition,

$$\vec{\nabla} \cdot \vec{A} = -\frac{1}{c^2} \frac{\partial}{\partial t} \phi, \tag{B4}$$

which now could be called the Lorenz relationship.

## Appendix C: Derivation of Maxwell's equations

Taking the derivative with respect to (unprimed) time of both sides of equation (2.17) and substituting into the wave equation (2.8a), we obtain

$$\vec{\nabla} \cdot \left( \vec{\nabla}\phi + \frac{\partial}{\partial t}\vec{A} \right) = -\frac{1}{\varepsilon_{\mathrm{v}}}\rho. \tag{C1}$$

Similarly, substituting the well-known identity $\nabla^2\vec{A} = \vec{\nabla}(\vec{\nabla}\cdot\vec{A}) - \vec{\nabla}\times(\vec{\nabla}\times\vec{A})$ and Eqn. (2.18) into (2.9a), we get

$$\vec{\nabla}\times(\vec{\nabla}\times\vec{A}) + \frac{1}{c^2}\frac{\partial}{\partial t}\left( \vec{\nabla}\phi + \frac{\partial}{\partial t}\vec{A} \right) = \mu_{\mathrm{v}}\vec{j}. \tag{C2}$$

We immediately recognize the terms in the parentheses,

$$\vec{E} = -\vec{\nabla}\Phi - \frac{\partial}{\partial t}\vec{A}, \tag{C3}$$

$$\vec{B} = \vec{\nabla}\times\vec{A}, \tag{C4}$$

as the usual definitions of the *electric* and *magnetic* fields evaluated at the *present* location, $\vec{r} = (x, y, z)$, and time. These two equations have also been derived previously from Maxwell's equations (Griffiths 1999, Jackson 1999) (in which case the latter are taken as postulates) or from the Helmholtz theorem (Jefimenko 1989).

Returning to Eqns. (C1) and (C2), the expressions for the fields immediately give

$$\vec{\nabla}\cdot\vec{E} = \frac{\rho}{\varepsilon_{\mathrm{v}}}, \tag{C5}$$

and

$$\vec{\nabla}\times\vec{B} = \frac{1}{c^2}\frac{\partial}{\partial t}\vec{E} + \mu_{\mathrm{v}}\vec{j}, \tag{C6}$$

which represent *Maxwell's first* and *fourth equations*, respectively.

*Maxwell's second equation* is easily derived. By taking the curl of both sides of Eqn. (C3) and recalling from vector calculus that the curl of the gradient is zero, we have

$$\vec{\nabla}\times\vec{E} = -\frac{\partial}{\partial t}\left( \nabla\times\vec{A} \right),$$

which, using Eqn. (3.4), becomes *Maxwell's second equation*, i.e.,

$$\vec{\nabla} \times \vec{E} = -\frac{\partial \vec{B}}{\partial t}. \tag{C7}$$

Similarly, *Maxwell's third equation* is easily obtained by applying the divergence operator to both sides of equation (C4) and recalling that the divergence of the curl is zero. Thus, we get

$$\nabla \cdot \vec{B} = 0, \tag{C8}$$

which is also known as "Gauss's law for magnetism."

In summary, Maxwell's first and fourth equations rely on the use of the wave equations for the scalar and vector potentials, the relation between them (i.e., the Lorentz 'condition'), and the use of two quantities that reduce the second order equations for potentials to first equations, which are recognized as the electric and magnetic fields. At the same time, the second and third Maxwell equations were derived solely from the definitions of the electric and magnetic fields.

## Appendix D: The relativistic forms of the potentials in the real space

Let us consider a point charge, $q$, moving in an arbitrary direction with instantaneous velocity, $\vec{v_c}(t)$, (see Fig. 3.1), and whose instantaneous position is denoted by $\vec{r'_c}(t)$. We write the charge density as

$$\rho\left(\vec{r'}, t\right) = q\delta^3\left[\vec{r'} - \vec{r'_c}(t)\right], \tag{D1}$$

which, substituted into Eqn. (3.14) and after using the sifting property of the delta function, leads to

$$\phi(\vec{r}, t) = \frac{q}{4\pi\varepsilon_v}\int_{-\infty}^{t} d\tau \frac{1}{|\vec{r} - \vec{r'_c}(\tau)|}\delta\left\{\tau - \left[t - \frac{|\vec{r} - \vec{r'_c}(\tau)|}{c}\right]\right\} - \frac{q}{4\pi\varepsilon_v}\int_{-\infty}^{t} d\tau \frac{1}{|\vec{r} - \vec{r'_c}(\tau)|}\delta\left\{\tau + \left[t - \frac{|\vec{r} - \vec{r'_c}(\tau)|}{c}\right]\right\}, \tag{D2}$$

To integrate equation (D2), we need to change the integration variable so that it becomes the same as the argument of the delta functions, i.e.,

$$t''_r \equiv \tau - \left[t - \frac{|\vec{r} - \vec{r'_c}(\tau)|}{c}\right], \tag{D3r}$$

and

$$t''_a \equiv \tau - \left[t + \frac{|\vec{r} - \vec{r'_c}(\tau)|}{c}\right]. \tag{D3a}$$

Also,

$$d\tau = \frac{dt''_r}{dt''_r/d\tau}, \tag{D4r}$$

and

$$d\tau = \frac{dt''_a}{dt''_a/d\tau}. \tag{D4a}$$

The temporal derivatives of $t''_r$ and $t''_a$ in these equations may be written, from Eqns. (D3r) and (D3a), as

$$\frac{dt''_r}{d\tau} = 1 + \frac{1}{c}\frac{d}{d\tau}\left|\vec{r} - \vec{r'_c}(\tau)\right| = 1 + \frac{1}{c}\frac{d}{d\tau}\left\{\left[\vec{r} - \vec{r'_c}(\tau)\right]\cdot\left[\vec{r} - \vec{r'_c}(\tau)\right]\right\}^{1/2} = 1 - \frac{\vec{R_c}(\tau)}{R_c(\tau)}\cdot\frac{\vec{v}(\tau)}{c}, \tag{D5r}$$

and

$$\frac{dt''_a}{d\tau} = 1 - \frac{1}{c}\frac{d}{d\tau}\left|\vec{r} - \vec{r'_c}(\tau)\right| = 1 - \frac{1}{c}\frac{d}{d\tau}\left\{\left[\vec{r} - \vec{r'_c}(\tau)\right]\cdot\left[\vec{r} - \vec{r'_c}(\tau)\right]\right\}^{1/2} = 1 + \frac{\vec{R_c}(\tau)}{R_c(\tau)}\cdot\frac{\vec{v}(\tau)}{c}, \tag{D5a}$$

respectively. Using these substitutions, Eqn. (D2) becomes

$$\phi(\vec{r}, t) = \frac{q}{4\pi\varepsilon_v}\int_{-\infty}^{R/c} dt''_r \frac{1}{|\vec{R_c}(\tau)|}\frac{1}{1 - \frac{\vec{R_c}(\tau)}{|\vec{R_c}(\tau)|}\cdot\frac{\vec{v}(\tau)}{c}}\delta(t''_r - 0) - \frac{q}{4\pi\varepsilon_v}\int_{-\infty}^{-R/c} dt''_a \frac{1}{|\vec{R_c}(\tau)|}\frac{1}{1 + \frac{\vec{R_c}(\tau)}{|\vec{R_c}(\tau)|}\cdot\frac{\vec{v}(\tau)}{c}}\delta(t''_a - 0). \tag{D6}$$

Using the sifting property of the delta function, the first integral in Eqn. (D6) becomes

$$\left.\frac{1}{\left|\overrightarrow{R_c}(\tau)\right|}\frac{1}{1-\frac{\overrightarrow{R_c}(\tau)}{\left|\overrightarrow{R_c}(\tau)\right|}\cdot\frac{\overrightarrow{v}(\tau)}{c}}\right|_{t''_r=0},$$

which, since $t''_r = 0$ for $\tau = t - \left|\vec{r} - \overrightarrow{r'}_c(\tau)\right|/c = t - \left|\overrightarrow{R_c}(\tau)\right|/c$, may be rewritten as

$$\frac{1}{\left|\overrightarrow{R_c}\left(t-\frac{R_c}{c}\right)\right|}\frac{1}{1-\frac{\overrightarrow{R_c}\left(t-\frac{R_c}{c}\right)}{\left|\overrightarrow{R_c}\left(t-\frac{R_c}{c}\right)\right|}\cdot\frac{\overrightarrow{v}\left(t-\frac{R_c}{c}\right)}{c}}.$$

The second integral in (D6) needs more careful examination. Since $t''_a = 0$ falls above the upper limit of integration [i.e., $\delta(t''_a) = 0$ within the entire integration range] except for $R_c = 0$, the second integral is equal to zero always except for $R_c = 0$, when it becomes

$$\left.\lim_{R_c\to 0}\frac{1}{\left|\overrightarrow{R_c}(\tau)\right|}\frac{1}{1+\frac{\overrightarrow{R_c}(\tau)}{\left|\overrightarrow{R_c}(\tau)\right|}\cdot\frac{\overrightarrow{v}(\tau)}{c}}\right|_{t''_a=0}=\lim_{R_c\to 0}\frac{1}{\left|\overrightarrow{R_c}\left(t+\frac{R_c}{c}\right)\right|}\frac{1}{1-\frac{\overrightarrow{R_c}\left(t+\frac{R_c}{c}\right)}{\left|\overrightarrow{R_c}\left(t+\frac{R_c}{c}\right)\right|}\cdot\frac{\overrightarrow{v}\left(t+\frac{R_c}{c}\right)}{c}}=\infty.$$

To incorporate both of these cases, the second integral may be written with the help of the notation

$$\delta_{R_c}=\begin{cases}0,\ \mathrm{R}>0\\\lim_{R_c\to 0}\frac{1}{\left|\overrightarrow{R_c}\left(t+\frac{R_c}{c}\right)\right|}\frac{1}{1-\frac{\overrightarrow{R_c}\left(t+\frac{R_c}{c}\right)}{\left|\overrightarrow{R_c}\left(t+\frac{R_c}{c}\right)\right|}\cdot\frac{\overrightarrow{v}\left(t+\frac{R_c}{c}\right)}{c}}=\infty,\ \mathrm{R}=0.\end{cases}$$

Using all these results in (D6), we may write

$$\phi=\frac{q}{4\pi\varepsilon_v\left|\overrightarrow{R_c}\left(t-\frac{R_c}{c}\right)\right|}\frac{1}{1-\widehat{R_c}\left(t-\frac{R_c}{c}\right)\cdot\overrightarrow{\beta_c}\left(t-\frac{R_c}{c}\right)}-\frac{q}{4\pi\varepsilon_v}\delta_{R_c}, \tag{D7}$$

where $\overrightarrow{R_c}\left(t-\frac{R_c}{c}\right)\equiv\vec{r}-\overrightarrow{r'}_c\left(t-\frac{R_c}{c}\right)$, $\widehat{R_c}\left(t-\frac{R_c}{c}\right)\equiv\frac{\overrightarrow{R_c}\left(t-\frac{R_c}{c}\right)}{R_c\left(t-\frac{R_c}{c}\right)}$, and $\overrightarrow{\beta_c}\left(t-\frac{R_c}{c}\right)\equiv\frac{\overrightarrow{v_c}\left(t-\frac{R_c}{c}\right)}{c}$.

Similarly, the current density may be written as

$$\vec{j}\left(\overrightarrow{r'},t\right)=q\frac{d\overrightarrow{r'_c}}{dt'}\delta^3\left[\overrightarrow{r'}-\overrightarrow{r'}_c(t)\right]=q\overrightarrow{v_c}(t)\delta^3\left[\overrightarrow{r'}-\overrightarrow{r'}_c(t)\right], \tag{D8}$$

which, substituted into Eqn. (3.20), leads to

$$\vec{A}(\vec{r},t)=\frac{\mu_v q}{4\pi}\int_{-\infty}^{t}d\tau\,\frac{1}{\left|\vec{r}-\overrightarrow{r'_c}(\tau)\right|}\overrightarrow{v_c}(\tau)\,\delta\left[\tau-\left(t-\frac{\left|\vec{r}-\overrightarrow{r'_c}(\tau)\right|}{c}\right)\right]-\frac{\mu_v q}{4\pi}\int_{-\infty}^{t}d\tau\,\frac{1}{\left|\vec{r}-\overrightarrow{r'_c}(\tau)\right|}\overrightarrow{v_c}(\tau)\,\delta\left[\tau-\left(t+\frac{\left|\vec{r}-\overrightarrow{r'_c}(\tau)\right|}{c}\right)\right].$$
$$\tag{D9}$$

Following similar arguments as in the case of the scalar potential, equation (D9) leads to:

$$\vec{A}=\frac{q\mu_v}{4\pi\left|\overrightarrow{R_c}\left(t-\frac{R_c}{c}\right)\right|}\frac{c\vec{\beta}\left(t-\frac{R_c}{c}\right)}{1-\widehat{R_c}\left(t-\frac{R_c}{c}\right)\cdot\overrightarrow{\beta_c}\left(t-\frac{R_c}{c}\right)}-\frac{q\mu_v c\vec{\beta}(t)}{4\pi}\delta_{R_c}, \tag{D10}$$

where, as above, $\overrightarrow{R_c}\left(t - \frac{R_c}{c}\right) \equiv \vec{r} - \overrightarrow{r'}_c(t - \frac{R_c}{c})$, $\widehat{R_c}(t - \frac{R_c}{c}) \equiv \frac{\overrightarrow{R_c}(t - \frac{R_c}{c})}{R_c(t - \frac{R_c}{c})}$, and $\overrightarrow{\beta_c}(t - \frac{R_c}{c}) \equiv \frac{\overrightarrow{v_d}(t - \frac{R_c}{c})}{c}$.

If the point charge in Fig. 1 is replaced by a point dipole whose moment $\vec{\mu}$ is time dependent, the polarization may be written as

$$\vec{P}\left(\overrightarrow{r'}, t\right) = \vec{\mu}(t)\delta^3\left[\overrightarrow{r'} - \overrightarrow{r'}_d(t)\right], \tag{D11}$$

where we replaced the subscript "c" in Fig. 1 by "d" to denote the dipole. When inserted into equation (3.25), Eqn. (D11) leads to the following expression for the electric Hertz potential:

$$\vec{\Pi}(\vec{r}, t) = \frac{q\mu_v}{4\pi}\int_{-\infty}^{t} d\tau \frac{1}{|\vec{r} - \overrightarrow{r'_d}(\tau)|}\vec{\mu}(\tau)\delta\left[\tau - \left(t - \frac{|\vec{r} - \overrightarrow{r'_d}(\tau)|}{c}\right)\right] - \frac{q\mu_v}{4\pi}\int_{-\infty}^{t} d\tau \frac{1}{|\vec{r} - \overrightarrow{r'_d}(\tau)|}\vec{\mu}(\tau)\delta\left[\tau - \left(t + \frac{|\vec{r} - \overrightarrow{r'_c}(\tau)|}{c}\right)\right]. \tag{D12}$$

Following the steps used in the derivation of Eqn. (D7), this equation becomes:

$$\vec{\Pi} = \frac{1}{4\pi\varepsilon_v\left|\overrightarrow{R_d}\left(t - \frac{R_d}{c}\right)\right|}\frac{\vec{\mu}\left(t - \frac{R_d}{c}\right)}{1 - \widehat{R_d}\left(t - \frac{R_d}{c}\right)\cdot\overrightarrow{\beta_d}\left(t - \frac{R_d}{c}\right)} - \frac{\vec{\mu}(t)}{4\pi\varepsilon_v}\delta_{R_d}, \tag{D13}$$

where $\overrightarrow{R_d}\left(t - \frac{R_d}{c}\right) \equiv \vec{r} - \overrightarrow{r'}_d\left(t - \frac{R_d}{c}\right)$, $\widehat{R_d}\left(t - \frac{R_d}{c}\right) \equiv \frac{\overrightarrow{R_d}\left(t - \frac{R_d}{c}\right)}{\left|\overrightarrow{R_d}\left(t - \frac{R_d}{c}\right)\right|}$, and $\overrightarrow{\beta_d}\left(t - \frac{R_d}{c}\right) \equiv \frac{\overrightarrow{v_d}\left(t - \frac{R_d}{c}\right)}{c}$.

## Appendix E: The k-space representation of the electric field of a point dipole

The compute the electric field of an oscillating dipole located at a fixed position $P'_d(x'_d, y'_d, z'_d)$, we first compute the scalar and vector potentials. When inserted into equation (2.13), equation (4.3),

$$\vec{\Pi}(\vec{r}, t) = \frac{c}{8\pi^3 \varepsilon_v} \iiint_{-\infty}^{\infty} d^3k \frac{1}{k} \int_{-\infty}^{t} d\tau \, \vec{\mu}(\tau) \sin[kc(t-\tau) - \vec{k} \cdot \overrightarrow{R_d}], \tag{4.3}$$

gives

$$\phi = -\vec{\nabla} \cdot \vec{\Pi} = -\frac{c}{8\pi^3 \varepsilon_v} \iiint_{-\infty}^{\infty} d^3k \frac{1}{k} \int_{-\infty}^{t} d\tau \left[\vec{\nabla} \cdot \vec{\mu}(\tau)\right] \sin[kc(t-\tau) - \vec{k} \cdot \overrightarrow{R_d}] -$$

$$\frac{c}{8\pi^3 \varepsilon_v} \iiint_{-\infty}^{\infty} d^3k \frac{1}{k} \int_{-\infty}^{t} d\tau \, \vec{\mu}(\tau) \cdot \vec{\nabla} \sin[kc(t-\tau) - \vec{k} \cdot \overrightarrow{R_d}]. \tag{E1}$$

But

$$\vec{\nabla} \cdot \vec{\mu}(\tau) = 0,$$

and

$$\vec{\nabla} \sin[kc(t-\tau) - \vec{k} \cdot \overrightarrow{R_d}] = -\cos[kc(t-\tau) - \vec{k} \cdot \overrightarrow{R_d}] \, \vec{\nabla}[\vec{k} \cdot \overrightarrow{R_d}] = -\cos[kc(t-\tau) - \vec{k} \cdot \overrightarrow{R_d}] \, [\vec{k} \times (\vec{\nabla} \times \overrightarrow{R_d}) + \overrightarrow{R_d} \times (\vec{\nabla} \times \vec{k}) + (\vec{k} \cdot \vec{\nabla})\overrightarrow{R_d} + (\overrightarrow{R_d} \cdot \vec{\nabla})\vec{k}] = -\cos[kc(t-\tau) + \vec{k} \cdot \overrightarrow{R_d}] \, [0 + 0 + (\vec{k} \cdot \vec{\nabla})\overrightarrow{R_d} + 0],$$

with

$$(\vec{k} \cdot \vec{\nabla})\overrightarrow{R_d} = (\vec{k} \cdot \vec{\nabla})\left(\vec{r} - \overrightarrow{r'_d}\right) = (\vec{k} \cdot \vec{\nabla})\vec{r} = \vec{k}.$$

Thus,

$$\vec{\nabla} \sin[kc(t-\tau) - \vec{k} \cdot \overrightarrow{R_d}] = -\vec{k} \cos[kc(t-\tau) - \vec{k} \cdot \overrightarrow{R_d}], \tag{E2}$$

and, therefore,

$$\phi = \frac{c}{8\pi^3 \varepsilon_v} \iiint_{-\infty}^{\infty} d^3k \int_{-\infty}^{t} d\tau \left[\hat{k} \cdot \vec{\mu}(\tau)\right] \cos[kc(t-\tau) - \vec{k} \cdot \overrightarrow{R_d}]. \tag{E3}$$

When inserted into equation (2.14), equation (4.3) gives

$$\vec{A} = \frac{1}{c^2} \frac{\partial \vec{\Pi}}{\partial t} = -\frac{c\mu_v}{8\pi^3} \vec{\mu}(t) \iiint_{-\infty}^{\infty} d^3k \frac{1}{k} \sin(\vec{k} \cdot \overrightarrow{R_d}) + \frac{1}{8\pi^3 \varepsilon_v} \iiint_{-\infty}^{\infty} d^3k \int_{-\infty}^{t} d\tau \, \vec{\mu}(\tau) \cos[kc(t-\tau) - \vec{k} \cdot \overrightarrow{R_d}]. \tag{E4}$$

If we were to switch to spherical coordinates in $\vec{k}$, we would replace $\vec{k} \cdot \overrightarrow{R_d}$ by $kR_d \cos\theta \equiv kR_d \cos\theta$, and the integral over k in the first term would become:

$$\iiint_{-\infty}^{\infty} d^3k \frac{1}{k} \sin(\vec{k} \cdot \overrightarrow{R_d}) = \int_{0}^{2\pi} d\varphi \int_{0}^{\pi} d\theta \sin\theta \int_{0}^{\infty} dk \, k \sin(kR_d \cos\theta),$$

which, after using the substitutions $u = \cos\theta$ and $du = -\sin\theta \, d\theta$ and integrating, would become:

$$2\pi \int_0^\infty dk\, k \int_1^{-1} du \sin(kR_d u) = 0.$$  (E5)

Thus, the first term in (E4) is zero, and we get:

$$\vec{A} = \frac{1}{8\pi^3 \varepsilon_v} \iiint_{-\infty}^{\infty} d^3k \int_{-\infty}^{t} d\tau\, \vec{\mu}(\tau) \cos\big[kc(t-\tau) - \vec{k}\cdot\vec{R_d}\big].$$  (E6)

The part of the electric field originating from the scalar potential Eqn. (E3) is

$$-\vec{\nabla}\phi = -\frac{c}{8\pi^3 \varepsilon_v} \iiint_{-\infty}^{\infty} d^3k \int_{-\infty}^{t} d\tau\, \big[\hat{k}\cdot\vec{\mu}(\tau)\big] \vec{\nabla} \cos\big[kc(t-\tau) - \vec{k}\cdot\vec{R_d}\big] = -\frac{c}{8\pi^3 \varepsilon_v} \iiint_{-\infty}^{\infty} d^3k \int_{-\infty}^{t} d\tau\, \big[\hat{k}\cdot$$

$$\vec{\mu}(\tau)\big] \vec{\nabla} \cos\big[kc(t-\tau) - \vec{k}\cdot\vec{R_d}\big].$$  (E7)

As with Eqn. (E2), it can be shown that

$$\vec{\nabla} \cos\big[kc(t-\tau) - \vec{k}\cdot\vec{R_d}\big] = \vec{k} \sin\big[kc(t-\tau) - \vec{k}\cdot\vec{R_d}\big],$$  (E8)

which, inserted in (E7), gives

$$-\vec{\nabla}\phi = -\frac{c}{8\pi^3 \varepsilon_v} \iiint_{-\infty}^{\infty} d^3k \int_{-\infty}^{t} d\tau\, \vec{k}\big[\hat{k}\cdot\vec{\mu}(\tau)\big] \sin\big[kc(t-\tau) - \vec{k}\cdot\vec{R_d}\big].$$  (E9)

The part of the electric field originating from the vector potential given by Eqn. (E6) is

$$-\frac{\partial}{\partial t}\vec{A} = \frac{1}{8\pi^3 \varepsilon_v} \vec{\mu}(t) \iiint_{-\infty}^{\infty} d^3k \cos\big(\vec{k}\cdot\vec{R_d}\big) - \frac{c}{8\pi^3 \varepsilon_v} \iiint_{-\infty}^{\infty} d^3k \int_{-\infty}^{t} d\tau\, k\, \vec{\mu}(\tau) \sin\big[kc(t-\tau) - \vec{k}\cdot\vec{R_d}\big].$$  (E10)

The total electric field obtained by summing up (E9) and (E10) is

$$\vec{E} = -\frac{c}{8\pi^3 \varepsilon_v} \iiint_{-\infty}^{\infty} d^3k \int_{-\infty}^{t} d\tau\, \vec{k}\big[\hat{k}\cdot\vec{\mu}(\tau)\big] \sin\big[kc(t-\tau) - \vec{k}\cdot\vec{R_d}\big] +$$

$$+ \frac{c}{8\pi^3 \varepsilon_v} \iiint_{-\infty}^{\infty} d^3k \int_{-\infty}^{t} d\tau\, k\, \vec{\mu}(\tau) \sin\big[kc(t-\tau) - \vec{k}\cdot\vec{R_d}\big] - \frac{1}{8\pi^3 \varepsilon_v} \vec{\mu}(t) \iiint_{-\infty}^{\infty} d^3k \cos\big(\vec{k}\cdot\vec{R_d}\big).$$  (E11)